\documentclass[final,aip,cha,author-year,preprint,amsmath,amssymb]{revtex4-1}

\usepackage{graphicx}
\usepackage{dcolumn}
\usepackage{bm}
\usepackage{xspace}
\usepackage{amsmath}
\begin{document}

\newcommand{\TE}[1]{\ensuremath{T_{X \rightarrow #1}}\xspace}
\newcommand{\infoflow}{\ensuremath{\mathbf{\varphi}_X}\xspace}
\newcommand{\infovel}{\ensuremath{\mathbf{\upsilon}_X}\xspace}


\title{Hidden Structures of Information Transport Underlying Spiral Wave Dynamics} 



\author{Hiroshi Ashikaga}
\email[]{hashika1@jhmi.edu}
\affiliation{Cardiac Arrhythmia Service, Division of Cardiology, Johns Hopkins University School of Medicine, 600 N Wolfe Street, Carnegie 568, Baltimore, MD 21287}
\author{Ryan G. James}
\email[]{rgjames@ucdavis.edu}
\affiliation{Copmlexity Sciences Center, Department of Physics, University of California, Davis, One Shields Avenue, Davis, CA 95616-8572}


\date{\today}

\begin{abstract}
A spiral wave is a macroscopic dynamic of excitable media that plays an important role in several distinct systems, including the Belousov-Zhabotinsky reaction, seizures in the brain, and lethal arrhythmia in the heart. Because spiral wave dynamics can exhibit a wide spectrum of behaviors, its precise quantification can be challenging. Here we present a hybrid geometric and information-theoretic approach to quantifying spiral wave dynamics. We demonstrate the effectiveness of our approach by applying it to numerical simulations of a two-dimensional excitable medium with different numbers and spatial patterns of spiral waves. We show that, by defining information flow over the excitable medium, hidden coherent structures emerge that effectively quantify the information transport underlying spiral wave dynamics. Most importantly, we find that some coherent structures become more clearly defined over a longer observation period. These findings validate our approach to quantitatively characterize spiral wave dynamics by focusing on information transport. Our approach is computationally efficient and is applicable to many excitable media of interest in distinct physical, chemical and biological systems. Our approach could ultimately contribute to an improved therapy of clinical conditions such as seizures and cardiac arrhythmia by identifying potential targets of interventional therapies.
\end{abstract}

\maketitle 

\begin{quotation}
We present a computationally efficient, hybrid geometric and information-theoretic approach to numerically quantifying the dynamics of spiral waves of excitable media. Here we view excitable media as a communication system where components share and/or transfer information. This view allows us to reduce the wide spectrum of spiral wave behaviors to information transport. The complex interaction between multiple spiral waves is incorporated in the trajectory of the transport. Our approach explains how information transport generates coherent structures underlying spiral wave dynamics. Application of this approach could have relevance to clinical conditions such as seizures and cardiac arrhythmia in which spiral waves play a major role.
\end{quotation}

\section{Introduction}

An excitable medium is a complex system composed of a number of components distributed in space that interact with each other. This interaction, typically through a reaction-diffusion process, leads to system behaviors at multiple scales. At the microscopic scale, the system behavior is characterized by excitation and relaxation of each component. This relatively simple micro-scale behavior, by creating a series of traveling waves, can generate a macro-scale behavior called a~\emph{spiral wave}~\citep{winfree1991varieties}. A spiral wave is a two-dimensional (2-D) traveling wave of excitation whose front is an involute spiral with increasing convex curvature toward the rotation center (`\emph{rotor}')~\citep{winfree1990vortex,jalife2004molecular}. The three-dimensional (3-D) equivalent of a spiral wave is called a \emph{scroll wave}~\citep{winfree1973scroll,wellner2002minimal,berenfeld2001shaping}.

Spiral waves are a macro-scale dynamic of excitable media that plays an important role in physical, chemical and biological systems, including the Belousov-Zhabotinsky reaction~\citep{winfree1984prehistory}, seizures in the brain~\citep{viventi2011flexible}, and lethal arrhythmia in the heart~\citep{gray1995mechanisms}. Spiral wave dynamics can exhibit a wide spectrum of behaviors even in the absence of extrinsic heterogeneity or anisotropy of the media or external perturbations. For example, a spiral wave can be stationary or drifting~\citep{davidenko1992stationary}. A spiral wave can exist alone~\citep{skanes1998spatiotemporal}, or coexist with other spiral waves~\citep{moe1964computer,allessie2010electropathological}. A single spiral wave can also break up into multiple spiral waves~\citep{fenton2002multiple}. Despite its strikingly memorable geometry, these behaviors make visual identification of spiral waves extremely challenging. This may have contributed to disappointing results of interventional catheter ablation therapy to target spiral waves to cure cardiac arrhythmia in clinical settings~\citep{buch2016long,mohanty2016impact}. Therefore, it is critically important to develop a method for quantitative assessment of spiral wave dynamics that is stable over the wide spectrum of behaviors over time. One of the traditional methods to quantify spiral wave dynamics is to use rotor dynamics as a surrogate~\citep{winfree1987time}. The rotor of a spiral wave can be defined as a phase singularity within a phase map~\citep{umapathy2010phase}. However, the relationship between rotors and spiral waves is not one-to-one: phase mapping tends to identify rotors where there is no spiral wave, whereas some spiral waves can have no rotors~\citep{kuramoto2003rotating}.

In this study, we present a hybrid geometric and information-theoretic approach to quantifying spiral wave dynamics as a macro-scale behavior of excitable media. Information theory is one of the predominant mathematical paradigms for studying a system's multi-scale structure~\citep{allen2014information}. In essence, we view excitable media as a communication system where components share and/or transfer information~\citep{ashikaga2015modelling} with their nearest neighbors, and reduce the wide spectrum of spiral wave behaviors to information transport which can be quantified as a surrogate for spiral wave dynamics. To define information transport in this system, we combine an information-theoretic measure of information flow~\citep{schreiber2000measuring} with the geometry of the excitable medium leading to an information velocity at each component. Similar to an electrical test particle, we define an information \emph{test particle} -- a point of negligible properties other than to be influenced by the field -- as a point that moves with the local information velocity. We then calculate the trajectory of information particles to derive coherent structures over the observation time period~\citep{mathur2007uncovering}. Our approach is computationally efficient in many excitable media of interest. We demonstrate the effectiveness of our approach by applying it to numerical examples of an excitable medium with different numbers and spatial patterns of spiral waves.

\section{Methods}

We performed the simulation and the data analysis using Matlab R2016a (Mathworks, Inc.).

\subsection{Model of an excitable medium}

We used a monodomain reaction-diffusion model that was originally derived by Fitzhugh~\citep{fitzhugh1961impulses} and Nagumo~\citep{nagumo1962active} as a simplification of the biophysically based Hodgkin-Huxley equations describing current carrying properties of nerve membranes~\citep{hodgkin1952quantitative}, which was later modified by Rogers and McCulloch~\citep{rogers1994collocation} to represent cardiac action potential. This model accurately reproduces several important properties of excitable media, including slowed conduction velocity, unidirectional block owing to wavefront curvature, and spiral waves.

\begin{align}
  \frac{\partial v}{\partial t} &= 0.26v(v-0.13)(1-v)-0.1vr+ I_{ex}+\nabla\cdot (D\nabla v)\\
  \frac{\partial r}{\partial t} &= 0.013(v-r)
  \label{eq:FHN02}
\end{align}

Here, $v$ is the excitation variable that can be identified with transmembrane potential, $r$ is the recovery variable, and $I_{ex}$ is the external current~\citep{pertsov1993spiral}. $D$ is the diffusion tensor, which is a diagonal matrix whose diagonal and off-diagonal elements are 1 and 0, respectively, to represent a 2-D isotropic system. We solved the model equations using a finite difference method for spatial derivatives and explicit Euler integration for time derivatives assuming Neumann boundary conditions.

\subsection{Simulation of spiral waves}

We simulated the system of a 2-D $120 \times 120$ isotropic lattice of components. The size of each component is 0.99 mm $\times$ 0.99 mm, therefore the size of the lattce is 11.9 cm $\times$ 11.9 cm. We simulated spiral waves by introducing 40 random sequential point stimulations in 40 random components of the lattice. In each component, we computed the time series with a time step of 0.063 msec, which was subsequently downsampled at a sampling frequency of 500 Hz to reflect realistic measurements in human clinical electrophysiology studies~\citep{fogoros2012electrophysiologic}. We excluded the stimulation period from the quantitative analysis. We included seven separate time series which included one through seven coexisting spiral waves at $t=$0 sec for a total duration of 60 seconds.

\subsection{Shannon entropy}

For each component, the time series of cardiac excitation was discretized to 1 when excited (during the action potential duration (APD) at 90\% repolarization, or $APD_{90}$) or 0 when resting. We treated each component on the lattice as a time-series process $X$, where at any observation time $t$ the process $X$ is either excited ($X_t=1$) or resting ($X_t=0$). Using this framework, one can compute the Shannon entropy $H$ (in bits) of each time-series process $X$:
\begin{equation}
  H(X) = -\sum_{x} p(x) \log_{2} p(x)
  \label{eq:entropy}
\end{equation}
where $p(x)$ denotes the probability density function of the time series generated by $X$, and we adopt the standard convention that $0 \cdot \log_2{0} = 0$. This quantifies the average information content of each component over the time history~\citep{ashikaga2015modelling}. To understand the impact of time history, the Shannon entropy and other information-theoretic indices were measured over various increasing time periods (e.g. 0-10 sec, 0-20 sec, ..., 0-60 sec).

\subsection{Information flow}

\begin{figure}
  \centering
  \includegraphics[width=\linewidth,trim={0 18cm 3.5cm 0},clip]{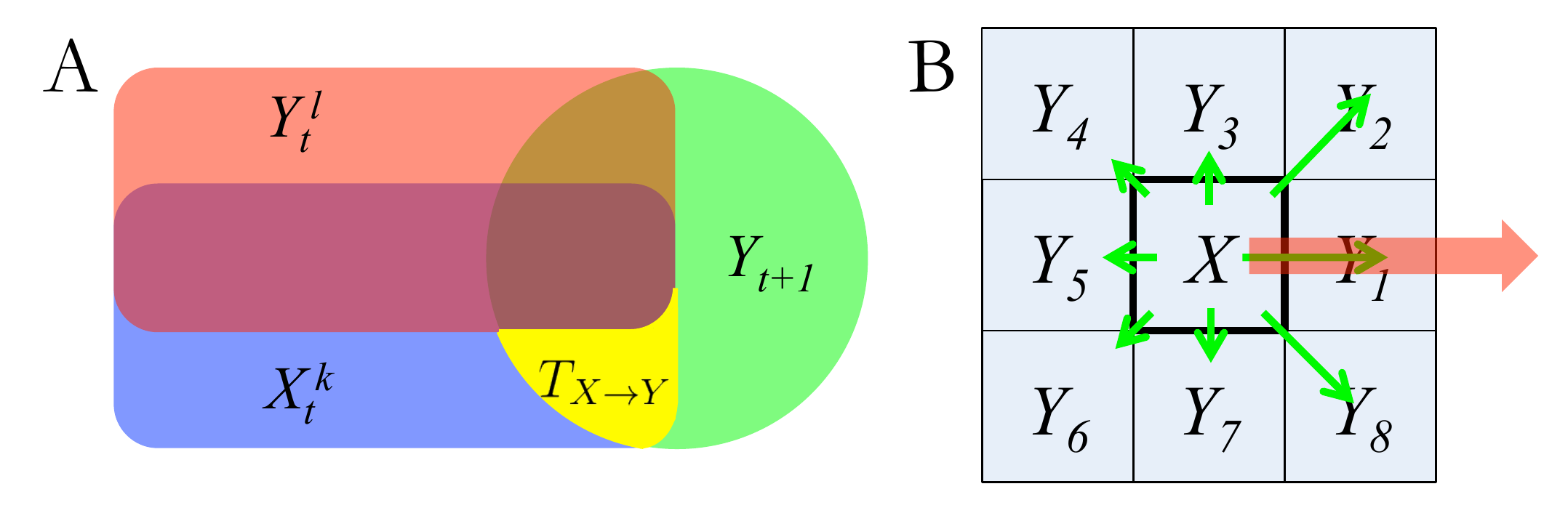}
  \caption{
    \textbf{Information flow in two dimensions.}
    \textit{A. Transfer entropy.} Transfer entropy \TE{Y} from a process $X$ (source) to another process $Y$ (destination) is the amount of uncertainty reduced in $Y_{t+1}$ by knowing $X_t$ given $Y_t$. $k$ and $l$ denote the length of time series in the processes $X$ and $Y$, respectively. The unit is [bits].
    \textit{B. Information flow.} Information flow of $X$ is the sum of transfer entropy vectors from $X$ (= one source) to the eight immediately adjacent neighbors (= eight destinations). The magnitude of information flow quantifies the information content transferred out of $X$ per unit time and is expressed in [bits/sec].
  }
  \label{fig:infodynamics}
\end{figure}

Transfer entropy~\citep{schreiber2000measuring} is a non-parametric statistic measuring the directed reduction in uncertainty in one time-series given another, generally interpreted as information transfer. In systems of unknown dynamics, the transfer entropy can result in values which should not be described as information flow~\citep{James2016}. The Fitzhugh-Nagumo dynamics used here, however, avoid such issues and so the transfer entropy is a reliable measure of information flow in this study. The transfer entropy from a process $X$ (source) to another process $Y$ (destination) is the amount of uncertainty reduced in future values of $Y$ by knowing the past values of $X$, given past values of $Y$ (Figure~\ref{fig:infodynamics}A):
\begin{align}
  \TE{Y}
  &= \sum p(y_{t+1},y^l_t,x^k_t) \log_2 \frac{p(y_{t+1}|y^l_t,x^k_t)}{p(y_{t+1}|y_t^l)} \\
  &= H(y_{t+1}|y^l_t) - H(y_{t+1}|y^l_t,x^k_t)
  ~,
  \label{eq:Txy}
\end{align}
where $k$ and $l$ denote the length of time series in the processes $X$ and $Y$, respectively:
\begin{align}
  x^k_t &= (x_t,x_{t-1},...,x_{t-k+1}) \\
  y^l_t &= (y_t,y_{t-1},...,y_{t-l+1})
  ~.
  \label{eq:xkyl}
\end{align}
Put another way, the transfer entropy quantifies how much better one can predict the value $y_{t+1}$ given both $x^k_t$ and $y^l_t$ over being give just $y^l_t$. In this study we define $k$ and $l$ such that $x^k_t$ and $y^l_t$ contain a unit time (= 1 sec) of the time-series preceding time $t$ ($k=l$). We used the discrete transfer entropy calculator of the Java Information Dynamics Toolkit (JIDT) to calculate transfer entropy~\citep{lizier2014jidt}.

In order to study the information dynamics of an excitable medium on a 2-D lattice, we consider \emph{information flow} between two adjacent sites $X$ and $Y$ to be a vector weighted by the transfer entropy \TE{Y}. This effectively combines the information-theoretic properties of the time-series with the geometry of the medium, resulting in spatiotemporal information dynamics. We define the 2-D information flow at $X$, \infoflow, as the sum of transfer entropy vectors from $X$ to its immediately adjacent neighbors. Here, we utilize the 8-site Moore neighborhood consisting of processes $Y_1, Y_2, \ldots, Y_8$ (see Figure~\ref{fig:infodynamics}B):

\begin{alignat}{6}
\mathbf{\varphi}_X =
  &&~& \TE{Y_4} \begin{pmatrix}
                  -1 \\
                  +1
                \end{pmatrix}
   &+& \TE{Y_3} \begin{pmatrix}
                   0 \\
                  +1
                \end{pmatrix}
   &+& \TE{Y_2} \begin{pmatrix}
                  +1 \\
                  +1
                \end{pmatrix} \nonumber \\
  &&+& \TE{Y_5} \begin{pmatrix}
                  -1 \\
                   0
                \end{pmatrix}
   &~&
   &+& \TE{Y_1} \begin{pmatrix}
                  +1 \\
                   0
                \end{pmatrix} \nonumber \\
  &&+& \TE{Y_6} \begin{pmatrix}
                  -1 \\
                  -1
                \end{pmatrix}
   &+& \TE{Y_7} \begin{pmatrix}
                   0 \\
                  -1
                \end{pmatrix}
   &+& \TE{Y_8} \begin{pmatrix}
                  +1 \\
                  -1
                \end{pmatrix}
\label{eq:outputvec}
\end{alignat}
The magnitude of information flow quantifies the information content transferred out of $X$ per unit time and is expressed in [bits/sec].

\subsection{Lagrangian coherent structures}

The displacement of information content per unit time, or \emph{information velocity} \infovel, at $X$ is given by:
\begin{align}
  \infovel = 0.99 \times \frac{4 \times (1 +\sqrt{2})}{8} \cdot \frac{\infoflow}{\left | \infoflow \right |} \mathrm{[mm/sec]}
  ~.
  \label{eq:velocity}
\end{align}
Advection of information through the lattice can be described in terms of the time-dependent velocity field $\mathbf{\upsilon}(\mathbf{x}, t)$:
\begin{align}
  \mathbf{\dot{x}}(\mathbf{x}_0, t_0, t) = \mathbf{\upsilon}(\mathbf{x}, t)
  ~,
  \label{eq:advection}
\end{align}
where $\mathbf{x}(\mathbf{x}_0, t_0, t)$ describes the trajectory of \emph{information particles} at time $t$, starting at position $\mathbf{x}_0$ at time $t_0$. An information particle is defined as a point that moves with the local velocity.

The deformation gradient tensor at position $\mathbf{x}_0$ between time $t_0$ and $t$ is given by:
\begin{align}
  \mathbf{F}^t_{t_0}(\mathbf{x}_0) =
  \begin{pmatrix}
    \frac{\partial \mathbf{x1}_t}{\partial \mathbf{x1}_{t_0}} &
    \frac{\partial \mathbf{x1}_t}{\partial \mathbf{x2}_{t_0}} \\
    \frac{\partial \mathbf{x2}_t}{\partial \mathbf{x1}_{t_0}} &
    \frac{\partial \mathbf{x2}_t}{\partial \mathbf{x2}_{t_0}}
    ~.
  \end{pmatrix}
  \label{eq:F}
\end{align}
The right Cauchy-Green deformation tensor at position $\mathbf{x}_0$ between time $t_0$ and $t$ is defined as \citep{truesdell2004non}:
\begin{align}
  \mathbf{C}^t_{t_0}(\mathbf{x}_0) = \mathbf{F}^t_{t_0}(\mathbf{x}_0)^T
                                     \mathbf{F}^t_{t_0}(\mathbf{x}_0)
  ~.
  \label{eq:C}
\end{align}
This symmtric tensor is positive definite due to the invertibility of $\mathbf{F}^t_{t_0}(\mathbf{x}_0)$. The eigenvalues $\lambda_i$ and associated eigenvectors $\xi_i$ of $\mathbf{C}$ satisfy the following properties:
\begin{align}
  \mathbf{C} \xi_i &= \lambda_i \xi_i \\
  \left |\xi_i \right | &= 1 \\
  i &= 1,2 \\
  0 &< \lambda_1 \leq \lambda_2 \\
  \xi_1 &\perp \xi_2
  \label{eq:eigen}
\end{align}

The finite-time Lyapunov exponent (FTLE) is defined as \citep{bollt2013applied}:
\begin{align}
  \mathbf{\Lambda}^t_{t_0}(\mathbf{x}_0) = \frac{1}{t-t_0} \ln{\sqrt{\lambda_2(\mathbf{x}_0)}}
  ~.
  \label{eq:ftle}
\end{align}
As its name might suggest, it is a finite time variant of the Lyapunov exponent~\citep{strogatz2014nonlinear} and quantifies the spatially- and temporally-localized divergence of trajectories. One can obtain the forward FTLE field by integrating the velocity forward in time from $t_0$ to $t$. Similarly, the backward FTLE field can be obtained by integrating the negative velocity field backward in time from $t$ to $t_0$. Note that the integration period (between $t_0$ and $t$) is the same for both forward and backward integration in this study.

The Lagrangian coherent structures (LCS) are defined as ridges, or lines of local maxima, of the FTLE field~\citep{shadden2005definition}. A sharp ridge is characterized by a high negative curvature, i.e. high negative eigenvalues $\mu$ of the Hessian matrix of the FTLE field $\nabla^2\mathbf{\Lambda}^t_{t_0}(\mathbf{x}_0)$. For points on the ridge, the gradient of the FTLE field $\nabla\mathbf{\Lambda}^t_{t_0}(\mathbf{x}_0)$ is tangent to the ridge line and perpendicular to the eigenvector $\eta$ corresponding to the smallest eigenvalues $\mu_{\mathrm{min}} < 0$ of the Hessian, which leads to the condition
\begin{align}
  \eta \cdot \nabla \mathbf{\Lambda}^t_{t_0}(\mathbf{x}_0) = 0
  ~.
  \label{eq:lcs}
\end{align}
Repelling and attracting LCS are derived from the forward and the backward FTLE fields, respectively. This framework can be expanded to a system on a 3-D lattice, though this study focuses on a 2-D system.

Lagrangian coherent structures serve as the organizing manifolds of a flow. Ideally, repelling LCS are somewhat akin to walls through which the flow does not traverse while attracting LCS are like channels through which flow is funneled. In real systems, such idealistic interpretations do not strictly hold but can be used to provide some intuition as to the behavior of the flow.

\section{Results}

\subsection{Behaviors of spiral waves}

\begin{figure}[!h]
  \centering
  \includegraphics[width=\linewidth,trim={0 6cm 1cm 0},clip]{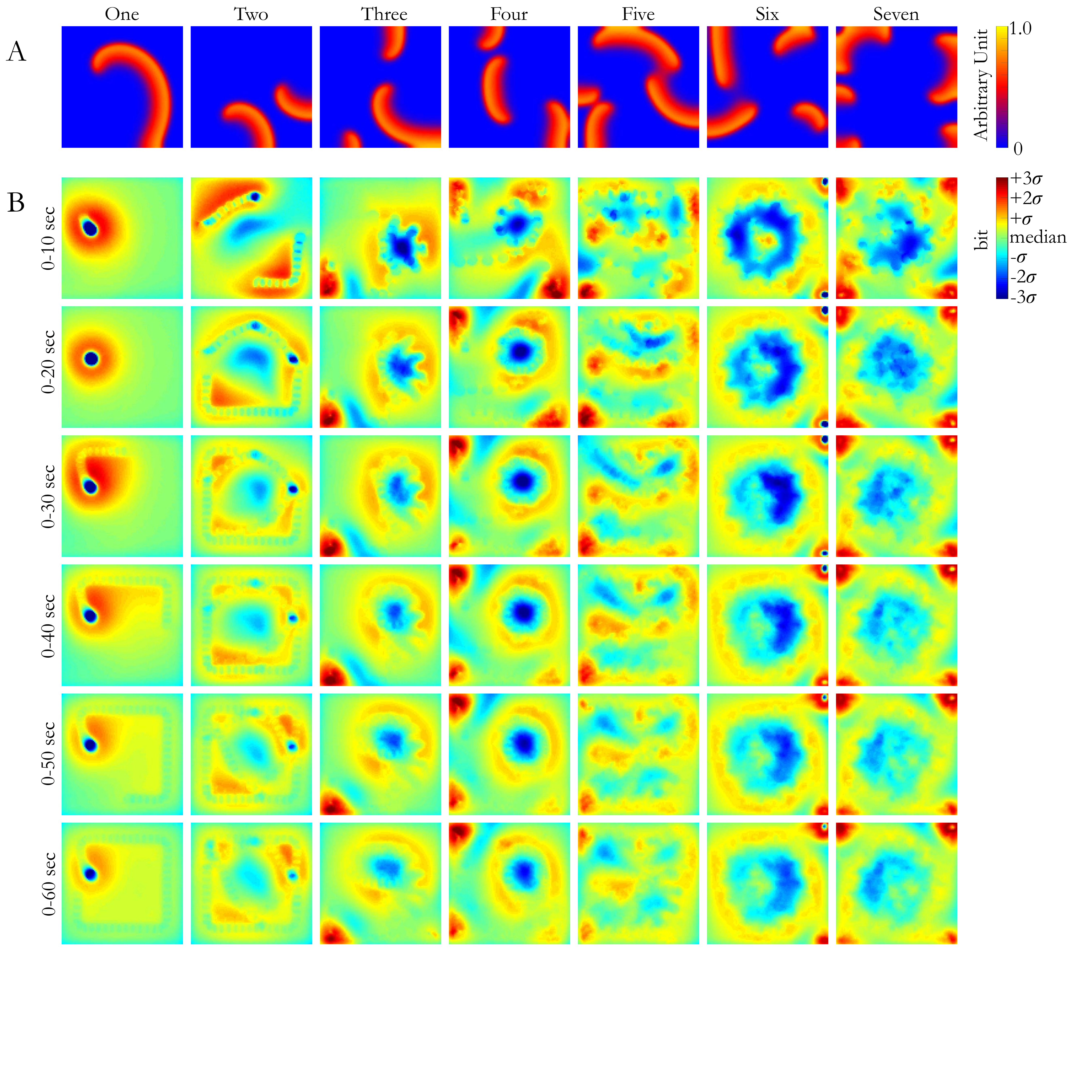}
  \caption{
    \textbf{A. Spiral waves.} A snapshot of one (far left panel) through seven (far right panel) spiral waves in a model of excitable media. The color represents a normalized transmembrane voltage between 0 and 1 in an arbitrary unit.
    \textbf{B. Shannon entropy of each component over various time periods.} The columns represent the number of spiral waves (one through seven) and the rows represent the time period during which the Shannon entropy is calculated (0-10 sec, 0-20 sec, 0-30 sec, 0-40 sec, 0-50 sec and 0-60 sec). The color scale is the median $\pm 3$ standard deviations ($\sigma$). The unit is bits.
  }
  \label{fig:entropy}
\end{figure}

We successfully reproduced spiral waves in the excitable medium. Figure~\ref{fig:entropy}A shows a snapshot of one (far left panel) through seven spiral waves (far right panel) in the lattice. In this study the system properties remain spatially homogeneous, isotropic and constant throughout the observation period, without external perturbations. However, the spiral waves exhibit a wide spectrum of spatial patterns and behaviors. For example, when only one spiral wave is present, the core of the spiral wave is initially stationary. It draws a near-circular trajectory in the left upper quadrant of the lattice while the spiral wave rotates counterclockwise, until it eventually drifts away from the original site (Supporting Movie 1). When there are two counter-rotating spiral waves, they drift by repelling each other (Supporting Movie 2). Of note, the rectangular trajectory of both the spiral waves represents an artificial impact determined by the boundary condition of the model. When there are three spiral waves, they also repel each other but exhibit a more complex behavior: one counterclockwise spiral wave is initially pushed away to the left lower corner of the lattice, while the remaining two clockwise spiral waves draw a clockwise circular co-orbit but $\pi$ out of phase (Supporting Movie 3). Similar spiral wave dynamics are observed when there are four (Supporting Movie 4), five (Supporting Movie 5), six (Supporting Movie 6) and seven rotors (Supporting Movie 7), where more than one spiral waves remain on the co-orbit and other spiral waves are pushed away into the corners of the 2-D lattice.

\subsection{Shannon entropy}

Figure~\ref{fig:entropy}B shows the Shannon entropy of each component over various time periods. When there is one spiral wave (far left column), the Shannon entropy shows a characteristic spatial profile of low values inside the core (shown in dark blue) in the left upper quadrant for the first 20 seconds (0-10 sec and 0-20 sec). This spatial profile represents a stationary, circular core during this period of time. The low Shannon entropy values within the core reflects no activity because the convex curvature of the spiral wave reaches a critical value~\citep{pandit2013rotors}. The core is surrounded by a doughnut-like region of high Shannon entropy values (shown in red) that reflects the high activity of the rotor, the wave front and the wave back around the core. Beyond this time period the core starts to drift, diminishing the relative magnitude of both the highest and the lowest values. For example, in the time period 0-60 sec, the core is still characterized by low values (shown in dark blue) surrounded by a doughnut-like region of high values (shown in red), but both the values are closer to the median compared with shorter time periods. These findings indicate that the Shannon entropy can identify the location of the core of a spiral wave if it is stationary for the time period of observation, but its utility diminishes when the spiral wave meanders.

The same observation is made for the case of two spiral waves (second column from far left) when spiral wave drift is particularly significant. For the first 10 sec (0-10 sec), both of the two cores drift, one from 9 o'clock to 12 o'clock, and the other from 3 o'clock to 6 o'clock, but the core regions still show low Shannon entropy values (shown in dark-light blue) that are surrounded by high values (shown in red). This contrast of high and low Shannon entropy values becomes less pronounced when the observation period becomes longer, until it is diluted in the background with the observation period of 60 sec (0-60 sec). A second low-entropy region at 3 o'clock manifests itself in the 0-20 and 0-30 time frames, and then fades in the longer observation windows. This is due to one of the cores pausing for a significant amount of time in that area before beginning to co-rotate again.

An interesting observation is made when there are three spiral waves (third column from far left). The core of the counterclockwise spiral wave that remains in the left lower corner of the lattice shows high Shannon entropy values (shown in red), lacking the central region with low Shannon entropy values that is observed in case of one spiral wave (far left column). This is because the rotor does not follow a circular but a more complex trajectory due to a constant interaction with the other two spiral waves. When the rotor trajectory is not circular, the core decreases in size or disappears completely, leaving only the region with high Shannon entropy values that reflect the high activity of the rotor, the wave front and the wave back. From a Shannon entropy perspective, the remaining two clockwise spiral waves that draw a clockwise circular co-orbit essentially behave as one large clockwise spiral wave. The central region between these two spiral waves has low Shannon entropy values, which is surrounded by a doughnut-like region of high values, making it appear similar to the case of a circular core of one spiral wave (far left column). This spatial profile persists for all time periods, from 0-10 sec to 0-60 sec. This finding suggests that a central region of low Shannon entropy values surrounded by a doughnut-like region of high values is not necessarily a signature profile of a spiral wave core.

A similar spatial profile of Shannon entropy is observed when there are four (center column), six (second from far right) and seven spiral waves (far right column). The spatial profile of Shannon entropy in the case of five spiral waves (third column from far right) is somewhat similar to the case of two spiral waves (second column from far left), since the spiral wave drift is significant. For the first 10 sec (0-10 sec), several localized regions of high (shown in red) and low (shown in blue) Shannon entropy values are identified. However, when the observation period becomes longer the regions of high and low Shannon entropy values are diluted in the background.

In summary:
\begin{enumerate}
  \item When the core is stationary and circular, the core shows low Shannon entropy surrounded by a doughnut-like region of high Shannon entropy,
  \item When the core is stationary and non-circular, the core shows high Shannon entropy,
  \item When the core drifts, the utility of Shannon entropy to characterize the core diminishes as the observation period becomes longer, and
  \item A region with low Shannon entropy surrounded by a doughnut-like region of high Shannon entropy does not necessarily represent the core of a spiral wave.
\end{enumerate}

\subsection{Instantaneous information flow}

\begin{figure}[!h]
  \centering
  \includegraphics[width=\linewidth,trim={0 12.5cm 5cm 0},clip]{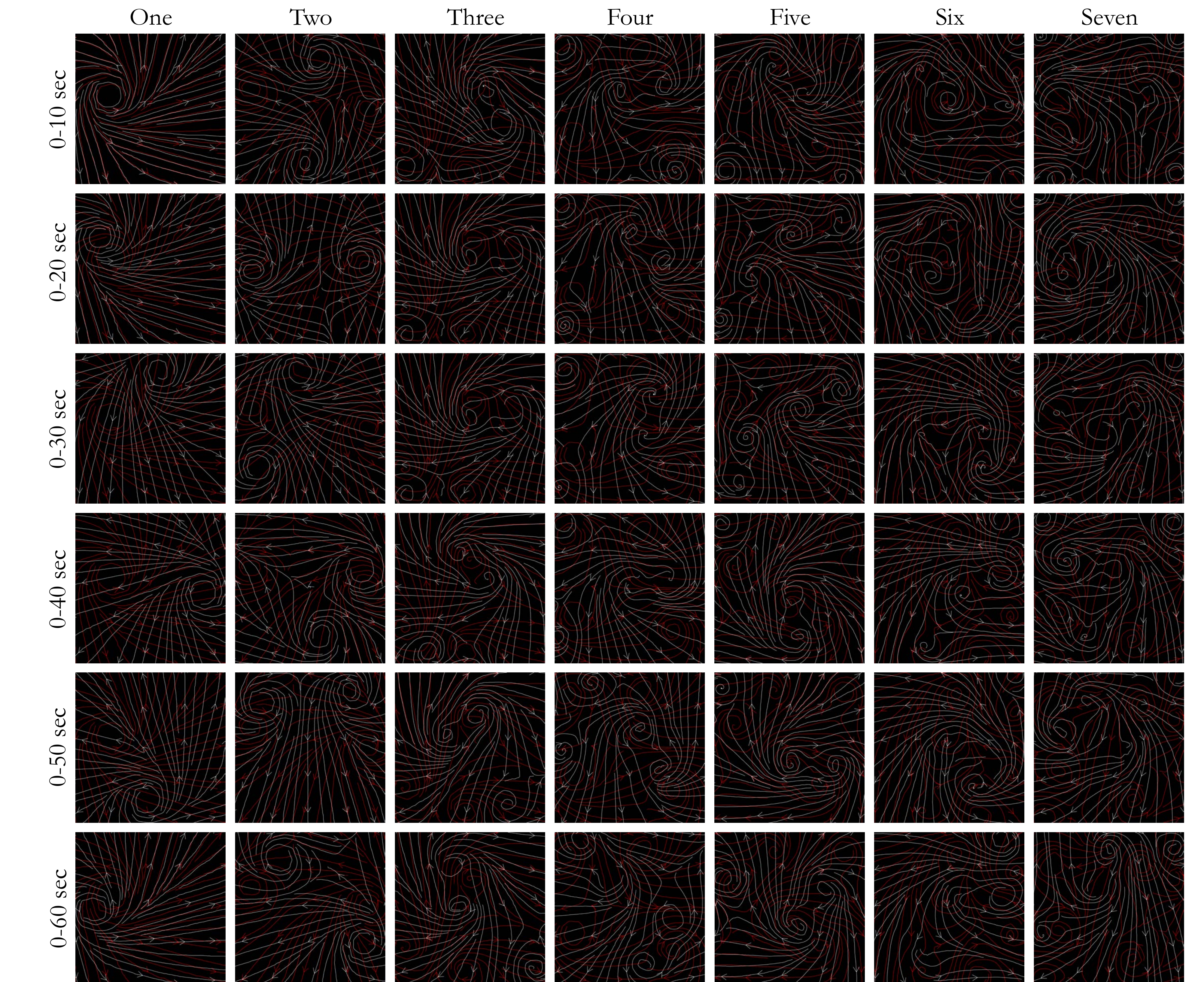}
  \caption{
    \textbf{Instantaneous information flow.} The columns represent the number of spiral waves (one through seven) and the rows represent the time period (e.g. 0-10 sec, 0-20 sec, 0-30 sec, 0-40 sec, 0-50 sec and 0-60 sec). In each panel, red and white streamlines represent the instantaneous information flow vector field at the beginning ($t=0$ sec) and the end of the observation period (e.g. $t=$10, 20, 30, 40, 50, and 60 sec), respectively.
  }
  \label{fig:streamlines}
\end{figure}

Figure~\ref{fig:streamlines} shows the instantaneous information flow vector field at the beginning ($t=0$ sec, red) and the end of each observation period (e.g. $t=$10, 20, 30, 40, 50, and 60 sec, white). When there is one spiral wave (far left column), the flow vector field at the beginning ($t=0$ sec, shown in red) and the end of the observation period at $t=10$ sec (top panel, shown in white) appears like an identical hurricane with its eye in the left upper quadrant, because the core remains stationary during this time period. The origin of information flow is the perimeter of the core where the rotors are, but no information is coming in or out of the core. This is because the spiral wave does not penetrate into the core, which remains quiescent for the entire observation period. The core location has drifted by the end of the observation period at $t=20$ sec (second panel from top, shown in white), and continues to drift over the subsequent time periods, although the hurricane-like appearance remains the same throughout all the observation periods.

When there are two spiral waves (second column from far left), the flow vector fields at the beginning and the end of each observation period are very different, because both spiral waves drift in the lattice while repelling each other. The streamlines indicate a saddle point of information flow between the two counter-rotating spiral waves, where the information flow converges in one direction and diverges in the other (e.g. top panel, shown in white). When there are three spiral waves, the streamlines of the two clockwise spiral waves on the same circular co-orbit converge, while the vectors between these spiral waves cancel each other. As a result, the two co-orbiting spiral waves act like a single clockwise spiral wave. The streamlines of the spiral wave that is pushed away in the corner show that they are slightly different between the beginning and the end of the observation period, indicating the presence of a non-circular core. A similar spatial profile of flow vector field is observed when there are four (center column), five (third column from far right), six (second column from far right) and seven spiral waves (far right column).

In summary:
\begin{enumerate}
  \item Instantaneous information flow vector field captures the time-dependent nature of information flow in spiral waves,
  \item The origin of information flow is the rotor of spiral waves,
  \item No information flow comes in or out of the core because the core remains quiescent,
  \item Multiple co-existing spiral waves can generate saddle points of information flow where the information flow converges in one direction and diverges in the other, and
  \item Multiple spiral waves can combine to act effectively as one spiral wave with a large core of low information flow because flow vectors of individual spiral waves cancel each other.
\end{enumerate}

\subsection{Total information flow over time}

\begin{figure}[!h]
  \centering
  \includegraphics[width=\linewidth,trim={0 12.5cm 1cm 0},clip]{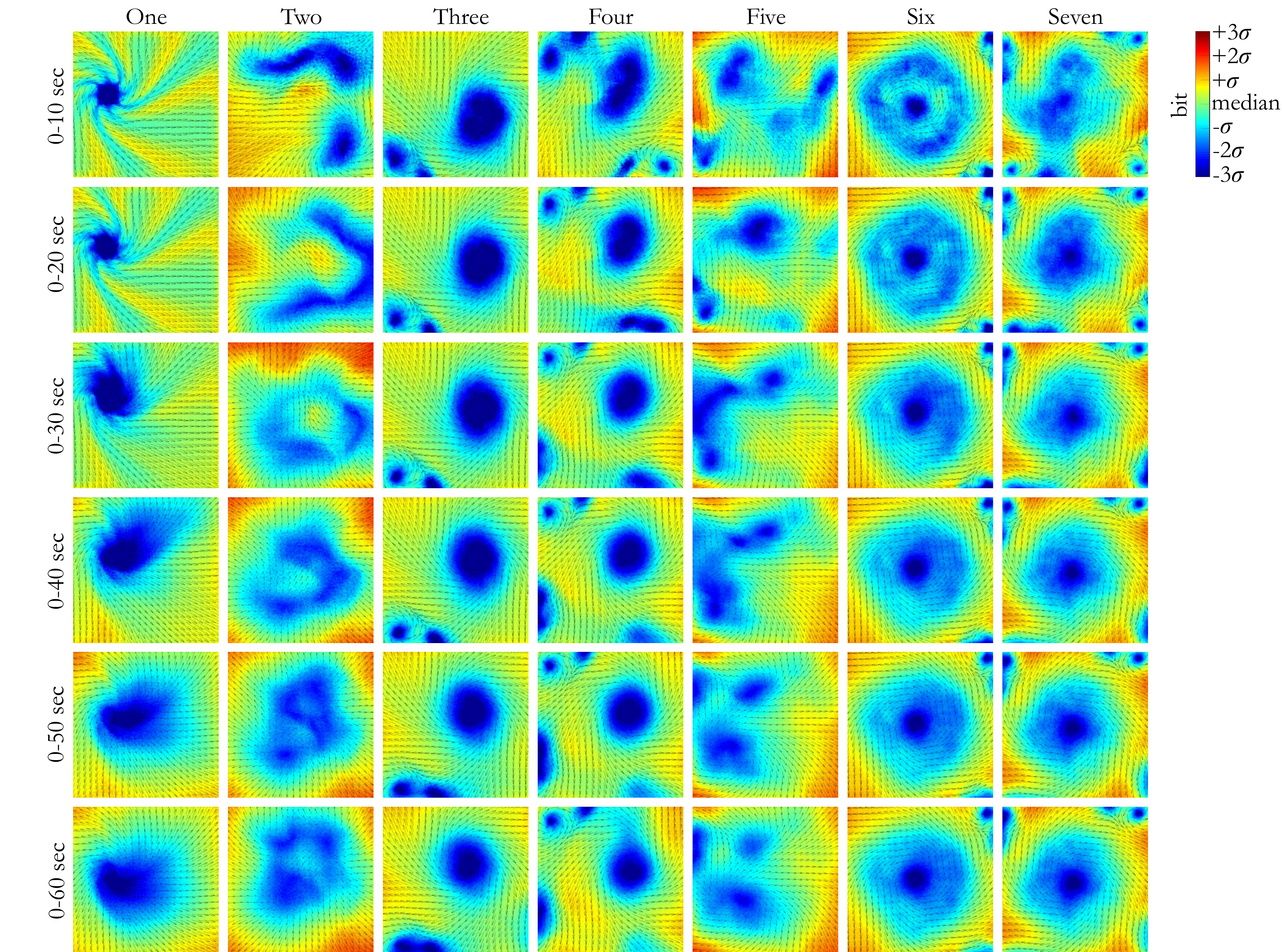}
  \caption{
    \textbf{Total information flow over time.} The columns represent the number of spiral waves (one through seven) and the rows represent the time period during which the sum of information flow vector is calculated (e.g. 0-10 sec, 0-20 sec, 0-30 sec, 0-40 sec, 0-50 sec and 0-60 sec). The color represents the magnitude of information output vector at each component, and the color scale is the median $\pm 3$ standard deviations ($\sigma$). The unit is bits.
  }
  \label{fig:info_out}
\end{figure}

Figure~\ref{fig:info_out} shows cumulative information flow of each component over various time periods (e.g. 0-10 sec, 0-20 sec, 0-30 sec, 0-40 sec, 0-50 sec and 0-60 sec). When there is one spiral wave (far left column), as expected from the instantaneous information flow vector field (Figure~\ref{fig:streamlines}), the spatial profile is characterized by low information output within the core. As the observation period lengthens, the area of low information output expands until it covers a large fraction of the center of the lattice (bottom panel). This is because the opposing information output vectors within the region circumscribed by the trajectory of the drifting spiral wave core cancel each other. The case of two (second column from far left) and five spiral waves (third column from far right) shows a similar phenomenon. When multiple spiral waves combine to act like one spiral wave, as in the case of three (third column from far left), four (center column), six (second column from far right) and seven spiral waves (far right column), the area of the combined core region (shown in dark blue) appears to remain relatively stable over various observation periods.

In summary:
\begin{enumerate}
  \item Total information flow over time is lowest in the core and becomes higher in the regions farther away from the core,
  \item Total information flow over time is low in the region circumscribed by the trajectory of the drifting spiral wave core, because the opposing information flow vectors within the region cancel each other, and
  \item As in Shannon entropy, a region with low total information flow is not necessarily the core of a spiral wave, because it could represent multiple spiral waves that combine to act like one spiral wave.
\end{enumerate}

\subsection{Information transport over time}

\begin{figure}[!h]
  \centering
  \includegraphics[width=\linewidth,trim={0 12.5cm 5cm 0},clip]{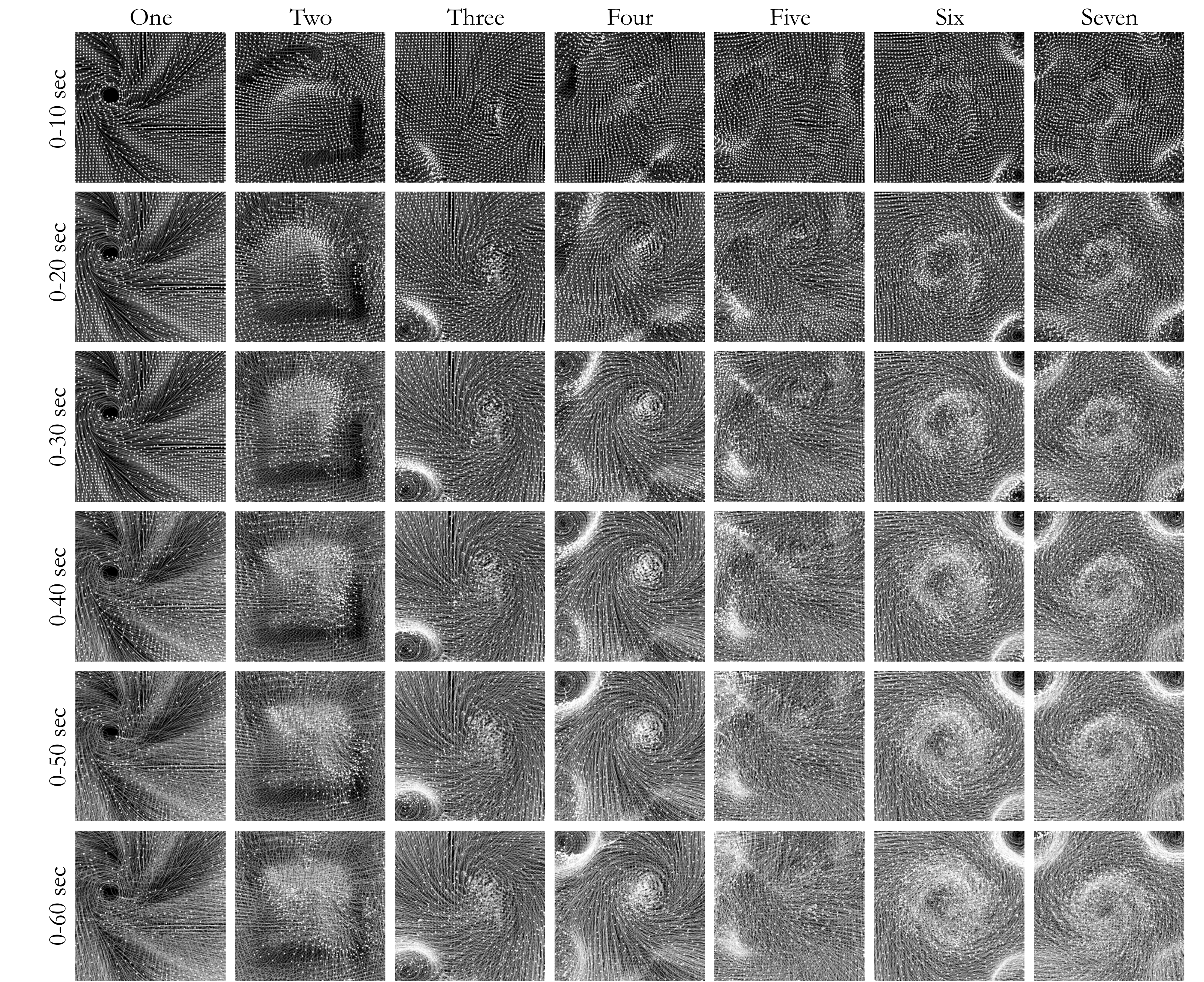}
  \caption{
    \textbf{Information transport.} The columns represent the number of spiral waves (one through seven) and the rows represent the time period during which the entire trajectory of information particles is calculated (e.g. 0-10 sec, 0-20 sec, 0-30 sec, 0-40 sec, 0-50 sec and 0-60 sec). In each panel, solid white lines represent the trajectory of information particles between the beginning ($t=$0 sec) and the end of each observation period (e.g. $t=$10, 20, 30, 40, 50, and 60 sec). Filled while circles represent the location of information particles at the end of each observation period. For simplicity, the trajectory of every four components in the lattice is shown.
  }
  \label{fig:traj}
\end{figure}

Figure~\ref{fig:traj} shows the trajectory of information particles over various time periods (e.g. 0-10 sec, 0-20 sec, 0-30 sec, 0-40 sec, 0-50 sec and 0-60 sec). When there is one spiral wave (far left column), the trajectory follows the streamlines of the instantaneous information flow vector field (Figure~\ref{fig:info_out}) for the first 20 seconds (0-10 sec and 0-20 sec), because the core is relatively stationary during this period of time. However, the density of information particles is not homogeneous. As the core begins to drift, the trajectory becomes more complex but the original structure of the spiral wave core remains in the left upper quadrant for the longest observation period (0-60 sec, bottom panel). By the end of 60 sec, the majority of information particles have traveled outside of the lattice, and the density of information particles in the lattice is sparse.

When two spiral waves are present (second column from far left), as in the case of one spiral wave, the information particles are dense in some regions and sparse in other regions. An interesting observation is that a large number of information particles are trapped within the region circumscribed by the trajectory of the two drifting spiral wave cores. These information particles form an obliquely linear structure that becomes clearer as the observation time period becomes longer (0-40 sec, 0-50 sec and 0-60 sec). As we shall see, this behavior is a clear example of an attracting LCS.

When there are three spiral waves, the information particles near the spiral wave trapped in the left lower corner create dense trajectories at the interface between the remaining two co-orbiting spiral waves and itself. The overall trajectory of information particles near the two co-orbiting spiral waves is similar to that of one spiral wave, but on close inspection oscillation of the trajectory is observed, which results from the complex interaction between the two co-orbiting spiral waves. A similar spatial profile is seen in the cases of four (center column), six (second column from far right) and seven spiral waves (far right column). When there are five spiral waves, the trajectory of information particles oscillates as well and is relatively more disorganized than other cases due to a lack of co-orbiting spiral waves. At the end of the longest time period ($t=60$ sec), information particles are scattered throughout the lattice with relatively homogeneous density.

In summary:
\begin{enumerate}
  \item When there is one spiral wave, information particles tend to travel away from the lattice and discrete patterns of trajectory disappear over time, and
  \item When multiple spiral waves are present, complex interaction between spiral waves tend to create discrete patterns of trajectory, which become clearer with longer observation periods.
\end{enumerate}

\subsection{Coherent structures over time}

\begin{figure}[!h]
  \centering
  \includegraphics[width=\linewidth,trim={0 12cm 1cm 0},clip]{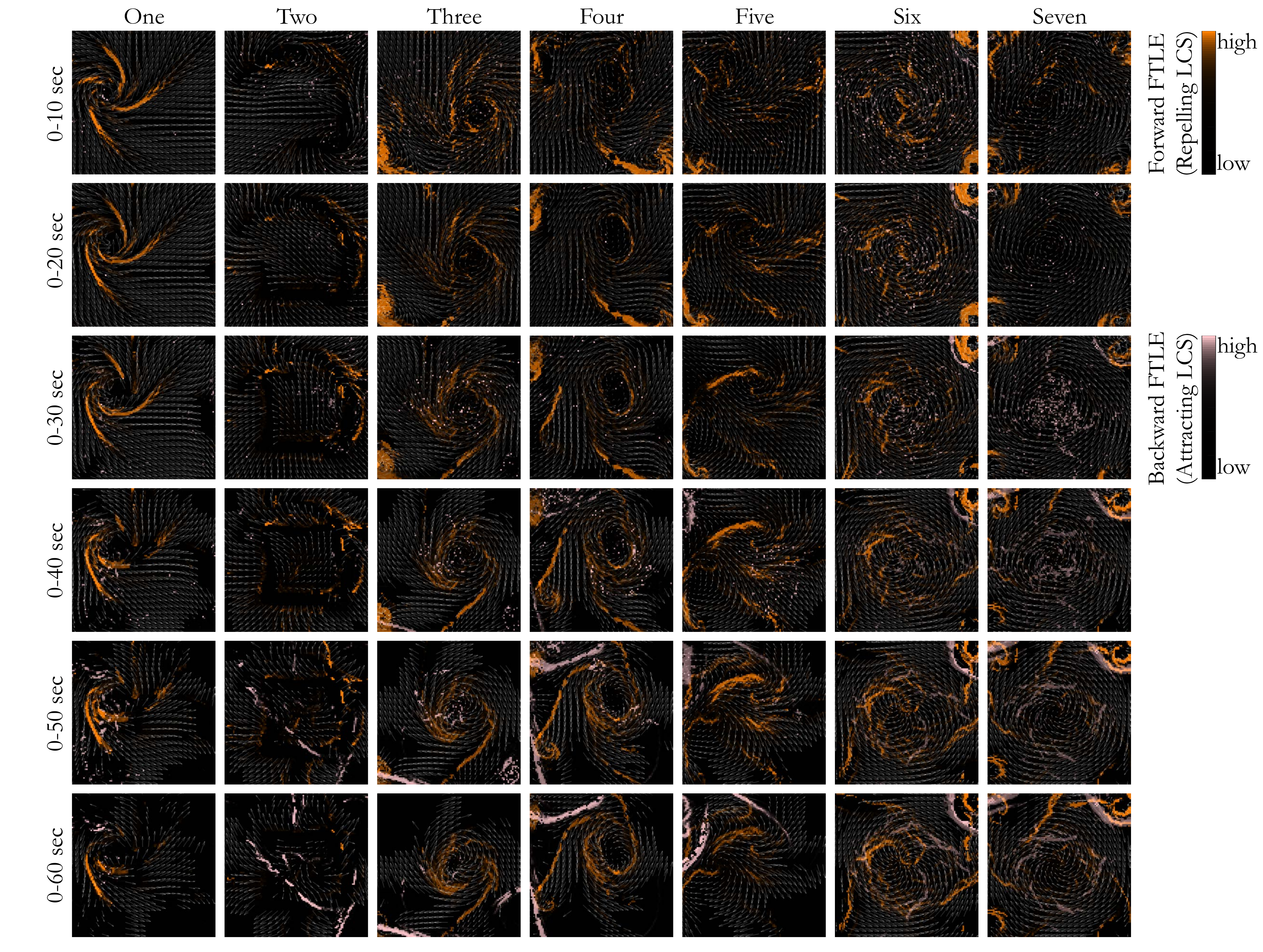}
  \caption{
    \textbf{Lagrangian coherent structures.} The columns represent the number of spiral waves (one through seven) and the rows represent the time period during which the entire trajectory of information particles is calculated (e.g. 0-10 sec, 0-20 sec, 0-30 sec, 0-40 sec, 0-50 sec and 0-60 sec). In each panel, orange and pink regions represent repelling and attracting Lagrangian coherent structures (LCS) over the observation period, respectively. The repelling LCS is the ridges of the forward finite-time Lyapunov exponents (FTLE), whereas the attracting LCS is the ridges of the backward FTLE.
  }
  \label{fig:lcs}
\end{figure}

Figure~\ref{fig:lcs} shows the Lagrangian coherent structures (LCS) of information flow over various time periods (e.g. 0-10 sec, 0-20 sec, 0-30 sec, 0-40 sec, 0-50 sec and 0-60 sec). When there is one spiral wave (far left column), as anticipated from the trajectory of information particles (Figure~\ref{fig:traj}), the repelling LCS (shown in orange) appears like a well demarcated hurricane with multiple arcs arising from the eye. As the observation period becomes longer, the hurricane-like repelling LCS decays until only a part of the arcs remain (0-60 sec). In contrast, the attracting LCS (shown in pink) is not apparent in shorter observation periods but begins to emerge as the observation period becomes longer, until it forms discrete linear structures in the left upper quadrant of the lattice (0-60 sec).

When two spiral waves are present (second column from far left), the repelling LCS no longer maintains the hurricane-like appearance. A transient arc of repelling LCS emerges at the right side of the lattice that outlines the trajectory of one spiral wave core (0-30 sec). However, the repelling LCS disappears almost entirely by the end of the longest observation period (0-60 sec, bottom panel). In contrast, linear attracting LCS only begins to appear with longer observation periods (0-50 sec and 0-60 sec). When three spiral waves are present (third column from far left), the repelling LCS that corresponds to the spiral wave in the left lower corner of the lattice is apparent in shorter observation periods, but it fades away as the observation period lengthens. Instead, the repelling LCS that represents the two co-orbiting spiral waves emerges over longer observation periods. In addition, several linear attracting LCS appear over longer observation periods corresponding to where trajectories accumulate at the interface between the stationary spiral and the two co-orbiting spirals. Similarly, when there are four (center column), five (third column from far right), six (second column from far right) and seven spiral waves (far right column), the repelling LCS of the spiral waves that are trapped in the corners of the lattice tends to disappear, while that of other spiral waves tend to evolve and emerge over longer time periods. The attracting LCS tends to appear more clearly in longer observations.

In summary:
\begin{enumerate}
  \item The geometry of both the repelling and the attracting LCS depends on the observation period,
  \item Some of the repelling LCS is defined clearly over a shorter observation period but disappear over a longer observation period, whereas other repelling LCS become more clearly defined over a longer observation period, and
  \item Attracting LCS is rarely seen in a shorter observation period but appears clearly over a longer observation period.
\end{enumerate}

\section{Discussion}

\subsection{Summary of main findings}

We find that Shannon entropy accurately identifies the location of a stationary core. When there is a single spiral wave, the core has minimal entropy surrounded by a donut of high entropy. When a spiral wave is stationary due to being pinned by the presence of other spirals, however, it may or may not have a low entropy core due to erratic jittering. However, the utility of Shannon entropy to describe spiral wave dynamics diminishes when the spiral wave drifts, more than one spiral waves co-exist, and/or the observation time is longer.

Additionally, we find that information flow originates from the rotor of spiral waves, and it captures complex interactions between multiple co-existing spiral waves such as saddle points of information flow. However, the utility of information flow to describe spiral wave dynamics diminishes over longer observation time because opposite information flow vectors of spiral waves can cancel each other, erase the history of information flow, and lead to overestimation of what appears to be the core area.

We further find that information transport, while preserving the history of information flow, can describe complex interactions between multiple co-existing spiral waves by creating discrete patterns of trajectory, which become clearer with longer observation periods. Lastly, we find that LCS effectively quantify the information transport underlying spiral wave dynamics. Most importantly, we find that some repelling and attracting LCS become more clearly defined over a longer observation period. These findings provide validity with our hybrid approach to quantitatively characterize spiral wave dynamics by focusing on information transport.

\subsection{Lagrangian perspective of information flow}

Our approach contains several innovative aspects. First, we extend the concept of transfer entropy~\citep{schreiber2000measuring} into a vector field on a 2-D lattice. This allows us to define information flow and velocity as vector quantities at each component, and to visualize information flow and velocity as time-dependent vector fields. Information flow of $X$ is the sum of transfer entropy vectors from $X$ to the eight immediately adjacent neighbors, and the magnitude of information flow quantifies the information content transferred out of $X$ per unit time [bits/sec]. Information velocity at $X$ is the displacement of information content per unit time [mm/sec]. These quantities are an ensemble of univariate analyses between one source and each of eight destinations of information transfer. A different potential approach to quantify an equivalent interaction is a multivariate analysis between one source and all of eight destinations. Such an approach, however, would produce undirected a scalar quantity, and thus would not achieve our objective to extend the transfer entropy into a vector quantity. In addition, part of the multivariate structure is captured by univariate interactions~\citep{lizier2011multivariate}, and the structure that is captured only by the multivariate interactions may be redundant~\citep{lizier2013towards} which can be quantified via the partial information decomposition~\citep{williams2010nonnegative,wibral2015partial}.

Second, we define a new concept of an information particle, as a point that moves with the local information velocity. This allows us to compute information transport in a Lagrangian perspective. For any vector field of unsteady flow, there are two fundamentally different frames of references~\citep{janicke2010measuring}. The first is an Eulerian perspective, where observation is made at components that are fixed in space. An Eulerian frame of reference is an intuitive approach, because the time series data derive directly from each component. For example, an Eulerian view is effective in providing site-specific information flow data at any point in time or over the entire time series. However, as shown in Figure \ref{fig:info_out}, information flow generated by spiral waves is so unsteady that opposite information flow vectors can cancel each other, erase the history of information flow, and lead to inaccurate characterization of dynamics, particularly over a longer observation time. Another vantage is a Lagrangian perspective, where we track hypothetical \emph{`particles'} that travel with the local velocity. A Lagrangian frame of reference is less intuitive but suited for quantifying a transport process~\citep{kai2009top}. For example, a Lagrangian view can define the origin and the fate of information particles over a specified time period.

Third, we uncover coherent structures underlying spiral wave dynamics by deriving the Lagrangian coherent structures (LCS) from the finite-time Lyapunov exponents (FTLE). This allows us to define organized patterns in the presence of inherently chaotic, time-dependent information flow. Even when full trajectory histories are available (Figure~\ref{fig:traj}), they are hard to interpret and the final transport pathways of particles are not necessarily obvious. In addition, due to the high sensitivity to initial conditions, the fate of individual particles is practically unpredictable. The concept of LCS was born in an attempt to describe the chaotic fluid dynamics by uncovering special surfaces of fluid trajectories that organize the rest of the flow into ordered patterns~\citep{haller2015lagrangian}. Repelling LCS serve as transport barriers, while attracting LCS indicate the backbones of mixing in moving particles~\citep{haller2011lagrangian}.

Our results clearly demonstrate that our approach can uncover the hidden structures underlying a macro-scale behavior of excitable media. We find that, although the behaviors of spiral wave are extremely dynamic over time, some LCS become more clearly defined over a longer observation period (Figure~\ref{fig:lcs}). This finding suggests that, despite a wide spectrum of apparent spiral wave behaviors, information transport underlying spiral waves follows organized patterns. Importantly, these patterns are not obvious from traditional voltage-based mapping of excitable media (Figure~\ref{fig:entropy}A). The repelling LCS of information transport indicates a surface barrier that separates indiviual information flow. In other words, the repelling LCS segment the lattice into smaller segments of information dynamics. In contrast, the attracting LCS of information transport represents a region of information mixing, which can be considered as a meeting point of information particles that originate from different spiral waves. Our approach provide a tool to quantitatively characterize a macro-scale behavior of excitable media by specifically focusing on information transport, thereby quantifying spiral wave dynamics.

\subsection{Clinical implications}

An advantage of our analysis lies in its generality of the model of excitable media. We used a simple model of an excitable medium so that the results from this model will be widely applicable. Therefore, our results allow one to gain general insights as to the coherent structures defined by information transport underlying spiral wave dynamics. Application of this approach could have relevance to clinical conditions such as seizures~\citep{viventi2011flexible} and cardiac arrhythmia~\citep{gray1995mechanisms} in which spiral waves play a major role. Our approach not only shows how information transport changes across different spiral wave dynamics, but also could ultimately contribute to an improved therapeutic intervention of these disorders by providing accurate diagnosis using metrics to quantify spiral wave dynamics. For example, the LCS that is relatively stable over time may be contributing to perpetuation of spiral waves. These LCS could be invasively intervened upon to promote \emph{pinning}~\citep{jimenez2012stationary} or \emph{anchoring}~\citep{rappel2015mechanisms} to stabilize the dynamics prior to elimination.

\subsection{Limitations}

There are two potential limitations to the study. First, we used a modified Fitzhugh-Nagumo model, which is a relatively simple model of excitable media, with a homogeneous and isotropic lattice. It is possible that our findings may not directly be extrapolated to a more realistic excitable medium with tissue heterogeneity and anisotropy. However, the information-theoretic metrics that we used in the study are independent of any specific trajectory of each spiral wave. Therefore, the concept of our approach is applicable to other excitable media. Second, we focused on the number of spiral waves in the excitable medium and thus did not consider the chirality of the spiral waves. Our results do not address how the chirality impacts the information transport and the coherent structures.

\subsection{Conclusions}

Information flow dynamics and the associated Lagrangian coherent structures effectively quantify the information transport underlying spiral wave dynamics. Most importantly, some repelling and attracting LCS become more clearly defined over a longer observation period. Our computationally efficient approach is applicable to many excitable media of interest in distinct physical, chemical and biological systems.

\section*{Supplementary material}

The observation is made between $t=0$ sec and $t=60$ sec in all the movies. The frame rate of the movies is intentionally made fast to minimize the file size.

\paragraph*{Supporting Movie 1.}
\label{S1_Movie}
{\bf One spiral wave.}

\paragraph*{Supporting Movie 2.}
\label{S2_Movie}
{\bf Two spiral waves.}

\paragraph*{Supporting Movie 3.}
\label{S3_Movie}
{\bf Three spiral waves.}

\paragraph*{Supporting Movie 4.}
\label{S4_Movie}
{\bf Four spiral waves.}

\paragraph*{Supporting Movie 5.}
\label{S5_Movie}
{\bf Five spiral waves.}

\paragraph*{Supporting Movie 6.}
\label{S6_Movie}
{\bf Six spiral waves.}

\paragraph*{Supporting Movie 7.}
\label{S7_Movie}
{\bf Seven spiral waves.}

\begin{acknowledgments}
This work was supported by a grant from the W. W. Smith Charitable Trust (to H.A.). RGJ was supported by, or in part by, the U. S. Army Research Laboratory and the U. S. Army Research Office under contracts W911NF-13-1-0390 and W911NF-13-1-0340. The authors thank John Mahoney for valuable input.
\end{acknowledgments}

\bibliography{flow_ref}

\end{document}